\documentclass[a4paper]{jpconf}
\usepackage{graphicx,amsmath,amssymb,subfigure}

\newcommand{\x}{\vec{x}}

\newcommand{\z}{\vec{z}}
\begin{document}
\title{Multi-frequency imaging of perfectly conducting cracks via boundary measurements}

\author{Won-Kwang Park$^1$ and Dominique Lesselier$^2$}

\address{$^1$Department of Mathematics, Kookmin University, Seoul, 136-702, Korea}
\address{$^2$D\'epartement de Recherche en \'Electromagn\'etisme - Laboratoire des Signaux et Syst\`{e}mes, UMR8506 (CNRS-Sup\'elec-Universit\'e Paris-Sud~11) 91192 Gif-sur-Yvette cedex, France}
\ead{parkwk@kookmin.ac.kr and dominique.lesselier@lss.supelec.fr}

\begin{abstract}
Imaging of perfectly conducting crack(s) in a 2-D homogeneous medium using boundary data is studied. Based on the singular structure of the Multi-Static Response (MSR) matrix whose elements are normalized by an adequate test function at several frequencies, an imaging functional is introduced and analyzed. A non-iterative imaging procedure is proposed. Numerical experiments from noisy synthetic data show that acceptable images of single and multiple cracks are obtained.
\end{abstract}

\section{Introduction}\label{sec:1}
One considers the inverse scattering problem of the retrieval of the shape of perfectly conducting cracks buried within a homogeneous medium. Data are scattered field  measured at the search domain boundary.

To our knowledge, inverse scattering with a single, perfectly conducting crack in 2-D satisfying a Dirichlet boundary condition has been studied in \cite{K}. In this reference, the retrieval algorithm was based on a Newton-type iterative scheme so that a good initial guess close to the unknown crack was needed to ensure convergence. But an iterative scheme generally requires a large amount of computational time, and it is hard to extend to the reconstruction of multiple cracks unless strong hypotheses are made beforehand.

So, non-iterative reconstruction algorithms have been developed, e.g., MUltiple SIgnal Classification (MUSIC)-type ones \cite{AKLP,PL1,PL2}, topological-derivative-based ones \cite{MKP,P3}. Recently, a multi-frequency algorithm has been proposed; related works are in \cite{AGKPS,P1,P2,PL3} and references. But in many cases, the analysis is not fully reliable, in particular since phenomena such as the unexpected appearance of ghost replicas cannot be explained.

Here, one is proposing an improved imaging algorithm for perfectly conducting cracks from scattered field data measured at the search domain boundary, and one analyzes its structure by designing a multi-frequency imaging functional weighted by each frequency. This involves that elements of the Multi-Static Response (MSR) matrix normalized by a test function can be represented in a suitable asymptotic form. Via careful analysis, one can infer a relationship between the proposed imaging functional and Bessel functions of integer order and explain why this imaging functional yields location and shape of the cracks.

The discussion goes as follows. In section \ref{sec:2}, the direct scattering problem is summarized and the asymptotic expansion formula introduced. In section \ref{sec:3} the multi-frequency-based imaging functional weighted by each frequency is put forth and its properties discussed. Numerical experiments are in section \ref{sec:4} for single and multiple cracks. Conclusion is in section \ref{sec:5}.

\section{Direct scattering problem and asymptotic expansion formula}\label{sec:2}
In this section, one considers 2-D electromagnetic scattering by a perfectly conducting crack, $\Gamma$, fully buried within a homogeneous domain $\Omega\subset\mathbb{R}^2$. One assumes that $\Gamma$ is a narrow, linear crack represented as follows
\begin{equation}\label{crackdescription}
\Gamma=\left\{\z:\z=(x,y),-\rho<x<\rho\right\},
\end{equation}
and that it does not touch the boundary $\partial\Omega$, i.e., $\Gamma\cap\partial\Omega=\O$.

Throughout this contribution, we remain with the Transverse Magnetic (TE) polarization case: at a given frequency $\omega_k\gg\rho$, let $u^{(n)}(\x;\omega_k)$ be the time-harmonic total field that satisfies
\begin{equation}\label{BVP}
\left\{\begin{array}{rcl}
  \Delta u^{(n)}(\x;\omega_k)+\omega_k^2 u^{(n)}(\x;\omega_k)=0 & \mbox{for} & \x\in\Omega\backslash\overline{\Gamma}, \\
  \noalign{\medskip}u^{(n)}(\x;\omega_k)=0 & \mbox{for} & \x\in\Gamma, \\
  \noalign{\medskip}\displaystyle\frac{\partial u^{(n)}(\x;\omega_k)}{\partial\vec{\nu}(\x)}=\frac{\partial e^{i\omega_k\vec{\theta}_n\cdot \x}}{\partial\vec{\nu}(\x)} & \mbox{for} & \x\in\partial\Omega.
\end{array}\right.
\end{equation}
Here, $\vec{\nu}$ is the outward unit normal vector to $\partial\Omega$ and $\{\vec{\theta}_n:n=1,2,\cdots,N\}$ are the $N$ vectors that describe the directions of incidence, assumed equi-distributed on the unit circle $\mathbb{S}^1$. In the same manner, $u_0^{(n)}(\x;\omega_k)$ satisfies (\ref{BVP}) without $\Gamma$. Then, the following asymptotic relationship of $u^{(n)}(\x;\omega_k)$ and $u_0^{(n)}(\x;\omega_k)$ holds uniformly on $\x\in\partial\Omega$ (see \cite{AKLP} for instance):
\begin{equation}\label{AsymptoticExpansionFormula}
  u^{(n)}(\x;\omega_k)-u_0^{(n)}(\x;\omega_k)=\frac{2\pi}{\ln(\rho/2)}u_0^{(n)}(\z;\omega_k)\mathcal{N}(\x,\z;\omega_k)+O\bigg(\frac{1}{|\ln\rho|^2}\bigg),
\end{equation}
where $0<\rho<2$ and $\mathcal{N}(\x,\z;\omega_k)$ is the Neumann function for $\Omega$,
\begin{equation}\label{NeumannFunction}
\left\{\begin{array}{rcl}
  \Delta \mathcal{N}(\x,\z;\omega_k)+\omega_k^2\mathcal{N}(\x,\z;\omega_k)=-\delta(\x,\z) & \mbox{for} & \x\in\Omega, \\
  \noalign{\medskip}\displaystyle\frac{\partial \mathcal{N}(\x,\z;\omega_k)}{\partial\vec{\nu}(\x)}=0 & \mbox{for} & \x\in\partial\Omega.
\end{array}\right.
\end{equation}

\section{Non-iterative multi-frequency imaging algorithm from MSR matrix}\label{sec:3}
One assumes that there exist $L$ different small cracks $\Gamma_l$ with same length $\rho$ centered at $\z_l$, $l=1,2,\cdots,L$, in $\Omega$. We apply the asymptotic formula (\ref{AsymptoticExpansionFormula}) to establish an imaging functional of the crack. For this, one uses the eigenvalue structure of the MSR matrix $\mathbb{M}(\omega_k)=\left(M_{nm}^{(k)}\right)_{n,m=1}^{N}$, where $M_{nm}^{(k)}$ denotes the normalized boundary measurement data:
\begin{equation}\label{ElementM}
  M_{mn}^{(k)}:=-\frac{1}{N}\int_{\partial\Omega}\bigg(u^{(n)}(\x;\omega_k)-u_0^{(n)}(\x;\omega_k)\bigg)\frac{\partial v^{(m)}(\x;\omega_k)}{\partial\vec{\nu}(\x)}dS(\x)\approx-\frac{2\pi}{N\ln(\rho/2)} \sum_{l=1}^{L}e^{i\omega_k(\vec{\theta}_n-\vec{\theta}_m)\cdot\z_l}.
\end{equation}
Here, one eliminates the residue term of (\ref{AsymptoticExpansionFormula}) and chooses a test function $v^{(m)}(\x;\omega_k)=e^{-i\omega_k\vec{\theta}_m\cdot\x}$.

Now, one designs an imaging functional. First, one performs the Singular Value Decomposition (SVD) of $\mathbb{M}(\omega_k)$, so that
\[\mathbb{M}(\omega_k)\approx\sum_{l=1}^{L}\vec{U}_l(\omega_k)\sigma_l(\omega_k)\overline{\vec{V}}_l^T(\omega_k),\]
where $\sigma_l(\omega_k)$ are (non-zero) singular values and where $\vec{U}_l(\omega_k)$ and $\vec{V}_l(\omega_k)$ are left and right singular vectors. Then, one introduces a weighted imaging functional at several discrete frequencies $\{\omega_k:k=1,2,\cdots,K\}$ as
\begin{equation}\label{ImagingFunction}
  \mathbb{E}(\x;K):=\sum_{k=1}^{K}\sum_{l=1}^{L}\omega_k\bigg(\overline{\vec{W}}(\x;\omega_k)\cdot\vec{U}_l(\omega_k)\bigg) \bigg(\overline{\vec{W}}(\x;\omega_k)\cdot\overline{\vec{V}}_l(\omega_k)\bigg),
\end{equation}
where
\[\vec{W}(\x;\omega_k)=\frac{\vec{D}(\x;\omega_k)}{|\vec{D}(\x;\omega_k)|}\quad\mbox{with}\quad
\vec{D}(\x;\omega_k)=\bigg(e^{i\omega_k\vec{\theta}_1\cdot\x},e^{i\omega_k\vec{\theta}_2\cdot\x},\cdots,e^{i\omega_k\vec{\theta}_N\cdot\x}\bigg)^T.\]
It should be noted that $\vec{W}(\x;\omega_k)$ satisfies the following properties (see \cite{AGKPS})
\begin{equation}\label{SimilarProperties}
  \vec{U}_l(\omega_k)\sim\vec{W}(\z_l;\omega_k)\quad\mbox{and}\quad
  \overline{\vec{V}}_l(\omega_k)\sim\vec{W}(\z_l;\omega_k).
\end{equation}
If one assumes that numbers $N$ and $K$ are large enough, applying (\ref{SimilarProperties}) to (\ref{ImagingFunction}) yields
\begin{align*}
  \mathbb{E}(\x;K)&\sim\sum_{l=1}^{L}\sum_{k=1}^{K}\omega_k\bigg(\sum_{s=1}^{N}e^{i\omega_k\vec{\theta}_s\cdot(\z_l-\x)}\bigg) \bigg(\sum_{t=1}^{N}e^{i\omega_k\vec{\theta}_t\cdot(\z_l-\x)}\bigg)\\
  &\approx\sum_{l=1}^{L}\int_{\omega_1}^{\omega_K}\omega\bigg(\int_{\mathbb{S}^1}e^{i\omega\vec{\theta}\cdot(\z_l-\x)}d\vec{\theta}\bigg)^2d\omega
  =\frac{1}{4\pi^2}\sum_{l=1}^{L}\int_{\omega_1}^{\omega_K}\omega J_0(\omega\vec{r}_l)^2d\omega,\quad\vec{r}_l:=|\z_l-\x|\\
  &=\frac{1}{4\pi^2}\sum_{l=1}^{L}\bigg[\frac{\omega_K^2}{2}\bigg(J_0(\omega_K\vec{r}_l)^2+J_1(\omega_K\vec{r}_l)^2\bigg) -\frac{\omega_1^2}{2}\bigg(J_0(\omega_1\vec{r}_l)^2+J_1(\omega_1\vec{r}_l)^2\bigg)\bigg],
\end{align*}
where $J_p(\cdot)$ denotes the Bessel function of order $p$ and of the first kind. A full derivation is to appear in an extended version of this contribution. Some properties of $\mathbb{E}(\x;K)$ are as follows:
\begin{itemize}
  \item $J_0(\omega|\z_l-\x|)^2+J_1(\omega|\z_l-\x|)^2$ reaches its maximum value at $\omega|\z_l-x|=0$. Hence, $|\mathbb{E}(\x;K)|$ plots are of large and small magnitudes at $\x=\z_l\in\Gamma_l$ and at $\x\ne\z_l\in\Gamma_l$, respectively.
  \item When $K$ is small, $|\mathbb{E}(\x;K)|$ plots exhibit replicas at local maxima and minima of $J_0$.
      %\[\mathbb{E}(\x;K)\sim\sum_{l=1}^{L}\sum_{k=1}^{K}\omega_kJ_0(\omega|\z_l-x|)^2\] so that
  \item $N$ must be large enough. It means that if one applies this algorithm with a small $N$, poor results follow. Moreover, by looking at (\ref{ElementM}), this analysis cannot be applied to the limited-view inverse scattering problem.
  \item Using frequencies high enough guarantees successful imaging performance.
  \item Based on the Statistical Hypothesis Testing \cite{AGKPS}, the proposed algorithm is robust vs. random noise when $K$ is large enough.
\end{itemize}

\section{Numerical experiments}\label{sec:4}
In this section, one introduces numerical examples for imaging small and long arc-like cracks. The operation frequency is taken of the form $\omega_k=2\pi/\lambda_k$; here $\lambda_k$, $k=1,2,\cdots,K(=10)$, is the given wavelength. The $\omega_k$ are equi-distributed in the interval $[\omega_1,\omega_K]$ with $\lambda_1=0.6$ and $\lambda_K=0.4$. A white Gaussian noise with $20$dB signal-to-noise ratio (SNR) is added to the unperturbed data.

\begin{figure}[!ht]
\begin{center}
\subfigure[$|\mathbb{E}(\x;10)|$ with $N=12$]{\includegraphics[width=0.32\textwidth]{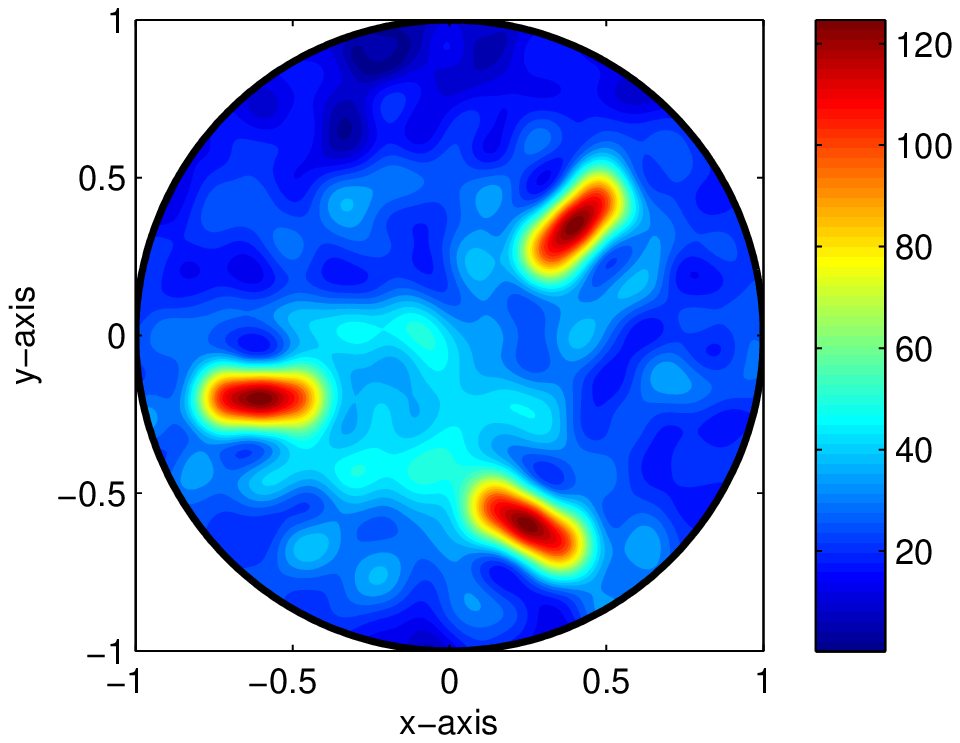}}
\subfigure[$|\mathbb{E}(\x;10)|$ with $N=32$]{\includegraphics[width=0.32\textwidth]{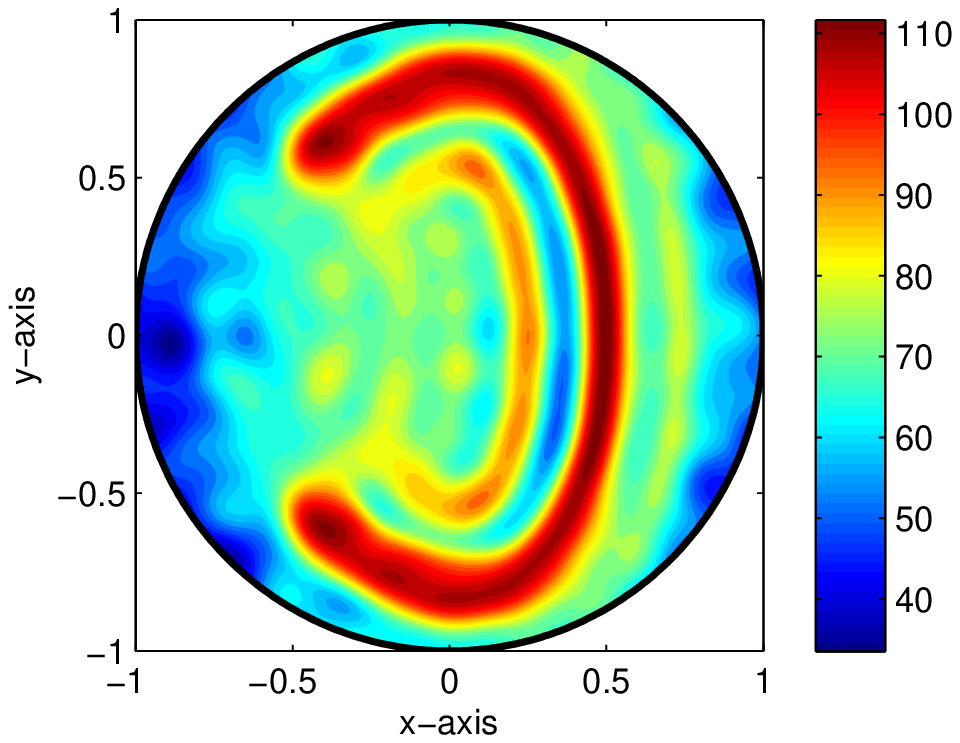}}
\subfigure[$|\mathbb{E}(\x;10)|$ with $N=36$]{\includegraphics[width=0.32\textwidth]{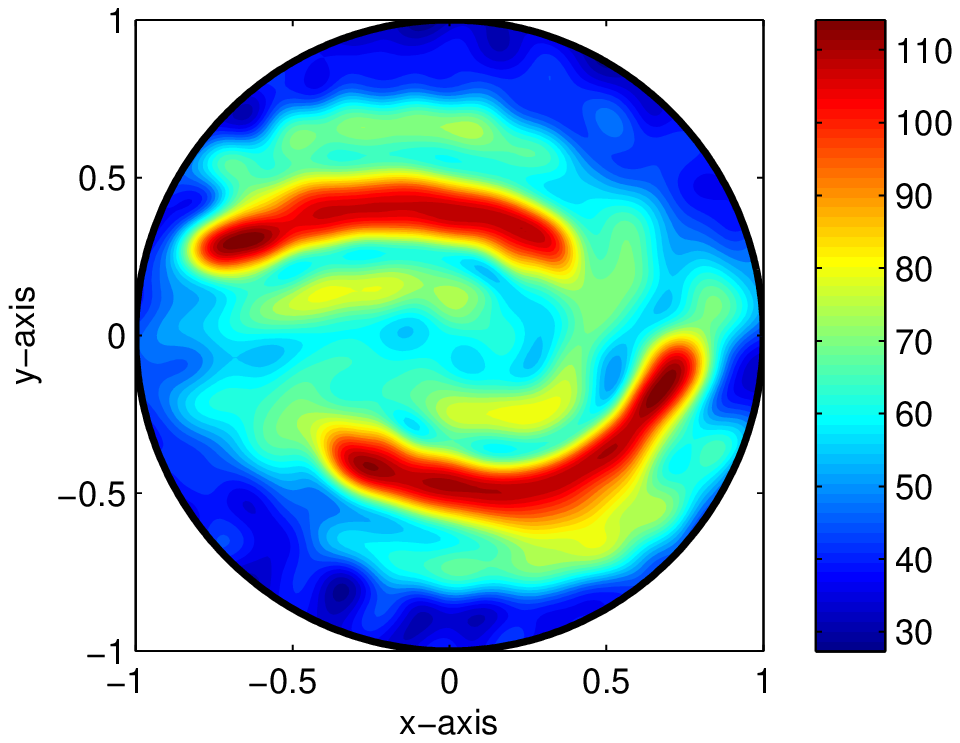}}
\end{center}
\caption{\label{ResultPlot}Map of $|\mathbb{E}(\x;K)|$ with $K=10$ for various cracks.}
\end{figure}

Figure \ref{ResultPlot} displays maps of $|\mathbb{E}(\x;10)|$ for three small cracks ({\it left}), a long single crack ({\it center}), and two long cracks close to one another ({\it right}). Notice that the analysis above is restricted to the small crack case, but it can be applied to extended (long) ones. Although some replicas appear since the chosen value of $K$ is rather small ($=10$), $|\mathbb{E}(\x;K)|$ plots exhibit maximum amplitudes at the location of the cracks so that one can recognize their shape. Taking them as initial guess could yield more accurate shapes via a Newton-type algorithm \cite{K} or a level-set method \cite{DL}.

\section{Conclusion}\label{sec:5}
Based on the structure of the singular vectors of the MSR matrix at several discrete frequencies of operation, an improved image function motivated from Kirchhoff migration has been proposed to image perfectly conducting cracks. The proposed imaging functional is related to Bessel functions of integer order. Numerical experiments with noisy data are discussed and they show the effectiveness of the corresponding imaging algorithm. The current contribution deals with a full-view inverse scattering problem. Application to a limited-view inverse problem is forthcoming work. Along the same line of thought, extension to  cracks with Neumann boundary condition should be an interesting subject.

\ack
This work was supported by Basic Science Research Program through the National Research Foundation of Korea (NRF) funded by the Ministry of Education, Science and Technology (No. 2012-0003207) and research program of Kookmin University in Korea.

\end{document}